\documentclass{statsoc}

\usepackage[a4paper]{geometry}
\usepackage{graphicx}
\usepackage[textwidth=8em,textsize=small]{todonotes}
\usepackage{amsmath}
\usepackage{natbib}
\usepackage{amsmath}
\usepackage{enumerate}
\usepackage{url}
\usepackage{amssymb}
\usepackage{bbding}
\usepackage{pifont}
\usepackage{wasysym}
\usepackage{lscape}

\usepackage{amsthm,amsfonts,amssymb}
\usepackage{adjustbox}
\usepackage{multirow}
\usepackage{hyperref}
\usepackage[figuresright]{rotating}
\usepackage{lscape}
\usepackage{array}
\newcolumntype{H}{>{\setbox0=\hbox\bgroup}c<{\egroup}@{}}
\usepackage{booktabs}
\usepackage{tabu}
\usepackage{threeparttable}

\def\defby{\stackrel{\mbox{\textrm{\tiny def}}}{=}}

\newtheorem{prop}{\bf Proposition}[section]

\def\beqr{\begin{eqnarray}}
	\def\eeqr{\end{eqnarray}}
\def\beqrs{\begin{eqnarray*}}
	\def\eeqrs{\end{eqnarray*}}
\def\bep{\begin{prop}}
	\def\eep{\end{prop}}

\makeatletter

\newcommand{\Rmnum}[1]{\expandafter\@slowromancap\romannumeral #1@}
\makeatother
\def\defby{\stackrel{\mbox{\textrm{\tiny def}}}{=}}

\def\beqr{\begin{eqnarray}}
	\def\eeqr{\end{eqnarray}}
\def\beqrs{\begin{eqnarray*}}
	\def\eeqrs{\end{eqnarray*}}
\def\bep{\begin{prop}}
	\def\eep{\end{prop}}

\title[Variable importance based interaction modeling]{Variable Importance Based Interaction Modeling with an Application on Initial Spread of  COVID-19 in China}
\author{Jianqiang Zhang}
\address{Center for Applied Statistics and School of Statistics, Renmin University of China, Beijing, China.}
\email{zhangjqs@163.com}
\author{Ze Chen}
\address{Center for Applied Statistics and School of Statistics, Renmin University of China, Beijing, China.}
\email{chze96@163.com}
\author{Yuhong Yang\thanks{\textbf{Correspondence} Yuhong Yang, School of Statistics, University of Minnesota, Minneapolis, USA. E-mail: yangx374@umn.edu}}
\address{School of Statistics, University of Minnesota, Minneapolis, USA.}
\email{yangx374@umn.edu}
\author[{Jianqiang Zhang et al.}]{Wangli Xu\thanks{\textbf{Correspondence} Wangli Xu, Center for Applied Statistics and School of Statistics, Renmin University of China, Beijing, China. E-mail: wlxu@ruc.edu.cn}}
\address{Center for Applied Statistics and School of Statistics, Renmin University of China, Beijing, China.}
\email{wlxu@ruc.edu.cn}

\begin{document}
\newpage
\begin{abstract}
Interaction selection for  linear regression models with both continuous and  categorical predictors is useful in many fields of modern science, yet very challenging when the number of predictors is relatively large.  Existing interaction selection methods focus on finding one
optimal model.  While attractive properties such as consistency and oracle property have been well established for such methods, they actually may  perform poorly in terms of stability for high-dimensional data, and they do not typically deal with categorical predictors. In this paper, we introduce a variable importance based interaction modeling (VIBIM) procedure for learning interactions in a linear regression model with both continuous and  categorical predictors. It delivers multiple strong candidate models with high stability and  interpretability. Simulation studies demonstrate its good finite sample performance. We apply the VIBIM procedure to a Corona Virus Disease 2019 (COVID-19) data used  in \cite{Tian:2020} and measure the effects of  relevant factors, including transmission control measures on the spread of COVID-19.  We show that the VIBIM approach  leads to better models in terms of interpretability,  stability, reliability and prediction.
\end{abstract}
\section{Introduction}
\label{sec:intro}

Interaction terms play an important role in many linear regression applications, e.g., applying user-item  interactions to identify higher level patterns in the recommender systems \citep{koren2009collaborative} and   identifying
interactions of single nucleotide polymorphisms (SNPs) for complex diseases  classification  \citep{Schwender:Ickstadt:2008}. Even if the number of predictors is moderate, in practice of linear regression, interactions are often barely considered or even completely ignored.  For instance, \cite{Tian:2020} did not consider interaction terms at all in studying the impact of transmission control measures on the initial spread of COVID-19. For a linear regression model with $p$ main effects,  the total number of  all pairwise interaction terms is  $\binom{p}{2}=O(p^2)$, which  increases drastically with $p$. Hence, identifying important interaction effects when $p$ is relatively large is very challenging.

In the existing literature, there are mainly two types of procedures for interaction selection, namely,  joint analysis \citep{Yuan:Zou:2009,Choi:Zhu:2010} and stage-wise analysis \citep{Bo:Liu:2014,Ning:Zhang:2014,HaoFengZhang_18}. The joint analysis approach selects the main and interaction effects simultaneously by making a global search over all possible models with interactions. The stage-wise analysis procedure  selects significant main effects only at the first stage, then identifies main effects and important interactions  among the reduced list of main effects. Although such procedures enjoy appealing asymptotic  properties, they may suffer from three  weaknesses, as will be elaborated later: (a) they are often unstable; (b) the procedures developed so far do not deal with categorical predictors; (c) they only select a single model among possibly many almost equally good  models. As a consequence, a novel approach is needed to overcome these shortcomings.

Regardless of which approach is used to select a final model, there is inherent model selection uncertainty. In fact, the problem of  instability associated with variable selection has been  well recognized in literature  (see, e.g., \cite{Draper:1995,Chatfield:1995,Yuan:Yang:2005,yu2017three}). Due to rapid increase in dimensionality incurred by interactions, the instability of interaction selection tends to be even higher. To measure the instability, there are several methods proposed in the existing literature, such as bootstrap resampling based measures \citep{Efron:1983:computer,breiman:1996:instability,buckland:1997:model}, sequential instability measure \citep{chen:2007:model} and variable selection deviation measure \citep{Nan:Yang:2014}.

To address high  variable selection uncertainty as revealed by the above measures, one may stabilize  a method (e.g., \cite{Buhlmann:2010:stability}, \cite{Lim:Yu:2016}), or rely on a method via minimizing  a variable selection uncertainty measure (e.g., \cite{yang:2017:VSD}). Such
methods have been shown to perform better. However, there is still an inherent drawback of the standard model selection approach, regardless of how it is done: conclusions based on the luckily selected model may not be reliable, since it presents only one story from the angle of the single model.
 When the data are not fully informative, as is often the case in moderate or high-dimensional data, it is more likely that multiple models are equally well supported by data at hand. In this scenario, as noted in \cite{Nan:Yang:2014}, any selection procedure will not yield a single model that significantly dominates all possible competitors, so that any selection method  chooses a lucky winner almost randomly among many more or less equally  promising models.

A well established  approach to address the issue of  ``winner takes all" is to use a model confidence set, which contains the true model with a given level of confidence; see \cite{Peter:2011}, \cite{Ferrari:Yang:2015}, \cite{li:2019:model}, \cite{zheng:2019:confidence} and references therein. From a practical point of view, however, the size of model confidence set is often too large to be  directly useful. In addition, models contained in a model confidence set may sometimes be drastically different in their compositions with few common variables. Thus, the  interpretability and applicability of the models in the model confidence set may not be necessarily satisfactory in our context.

In this paper, we propose a methodology to deal with all of the three weaknesses at the same time.
To address the model selection instability issue, we take advantage of model averaging that mitigates model selection uncertainty by weighting estimators across some models. There are various model averaging approaches proposed, e.g., Bayesian model averaging \citep{Hoeting:Madigan:Raftery:Volinsky:1999}, adaptive regression by mixing (ARM) \citep{Yang:2001}, model averaging based on a local asymptotic framework \citep{Claeskens:2003:Frequentist}, model averaging with estimator’s risk characteristics \citep{Leung:Barron:2006}, Mallows model averaging \citep{Hansen:2007} and parsimonious model averaging \citep{Zhang:Zou:Liang:Carroll:2020}.

Note that model averaging does not perform variable selection. However, model averaging weights  can be used in this regard. For instance, \cite{yang:2017:VSD} showed that such weighting based methods for variable selection become more stable. Thus,
combining the advantages of model averaging and variable importance, we develop a variable importance based interaction modeling (VIBIM) procedure using the sparsity oriented importance learning (SOIL) proposed by \cite{Ye:Yi:Yang:2018}. The detailed introduction of the VIBIM procedure is presented in Section 2.

Moreover, SOIL can be applied to  models with categorical variables and its theoretical properties are established in \cite{Chen:Zhang:Xu:Yang:2021}. Thus, VIBIM  overcomes the second weakness of the current two types of interaction selection procedures mentioned earlier.
Further, instead of glorifying a single selected model, VIBIM delivers multiple candidate models with high stability and  interpretability.

We evaluate the performance of VIBIM  by simulation studies. The results indicate the superiority of VIBIM  compared with existing approaches.
We apply   VIBIM to a data set on the initial spread of COVID-19 in China, pioneered in \cite{Tian:2020} with the goal to measure the effects of  relevant factors, including transmission control measures on the spread of the disease. We identify three major weaknesses of the published work in {\it Science}, and show that the VIBIM approach  leads to better models in terms of interpretability,  stability, reliability and prediction. In addition, we  correct a puzzling sign of an important covariate reported in \cite{Tian:2020}.

The remainder of the paper is organized as follows. In Section 2, the proposed VIBIM methodology  is described in details. Simulation analysis is carried out in Section 3 to  illustrate
the finite sample performance of the proposed procedure.  In Section 4,  the COVID-19 data and methods---linear and Poisson regression models are introduced. Section 5 applies the proposed procedure to investigate the effects of the relevant factors, including transmission control measures on the spread of COVID-19 with interpretation and discussions. Concluding remarks are offered in Section 6. Additional analysis results are provided in the Supplementary Material.
\section{Methodology}
\label{sec:methodology}
\subsection{Preliminaries}
\label{sec:preliminaries}
For a linear regression model with both categorical and continuous predictors, without loss of generality, we assume that, among the $p$ predictors  $\{X_1, \ldots, X_{p}\}$, the predictors $\{X_1, \ldots, X_q\}$ are categorical, while  $\{X_{q+1}, \ldots, X_p\}$ are continuous. The categorical levels of $\{X_1, \ldots, X_q\}$ are denoted by $\{J_1, \ldots, J_q\}$ respectively. For each categorical variable $X_i$, we define dummy variables $X_{i,j}$ pertaining to
the $j$th categorical level for $j=1,  \ldots, J_i-1$, and put $X_{\mathcal{I}_{i}}=(X_{i,1},\ldots,X_{i,J_i-1})^{\mathrm{T}}$ with $\mathcal{I}_{i}\defby\{(i,1),\ldots,(i,J_i-1)\}$ in the regression. In a similar fashion, put $X_{\mathcal{I}_{i}}=X_i$ with $\mathcal{I}_{i}\defby\{i\}$ for each continuous predictor $X_i$. A linear regression model with categorical and continuous predictors is written as
\beqr\label{linear1}
Y=\beta_0+\sum_{i=1}^p X_{\mathcal{I}_{i}}^{\mathrm{T}}  \beta_{\mathcal{I}_{i}}  +\epsilon.
\eeqr
Here, $\beta_{\mathcal{I}_{i}}=( \beta_{i,1},\ldots,\beta_{i,{J_i-1}})^{\mathrm{T}}$ for $i=1, \ldots, q$ and
$\beta_{\mathcal{I}_{i}}=\beta_i$ for $i=q+1, \ldots, p$. The error $\epsilon$ is assumed to be normally distributed with mean zero and variance $\sigma^2$. The total number of predictors in model \eqref{linear1} is $p^\ast=\sum_{i=1}^{q}J_i+p-2q$. For $N>1$, let $\mathcal{M}\defby\{M_i, i=1,\ldots,N\}$ be a candidate model set, where $M_i=\bigcup_{j\in \mathcal{A}_i}\mathcal{I}_{j}$, $\mathcal{A}_i \subseteq \{1,\ldots,p\}$.
\subsection{A new interaction selection procedure}
\label{sec:procedure}
To address the high instability of model selection approaches associated with a single selection criterion, one constructive technique is to take advantage of a sound variable importance (VI) measure.
As an effective tool,  VI can be used to arrive at more stable results.  To be specific, we consider
interactions among variables with reasonably large VI values. We propose to rank all the main effects and interactions according to VI values in a descending order and sort out top nested models. In this way, VI offers  most plausible models.
Various methods are available for evaluating variable importance. For instance, RFI1 and RFI2 investigated by \cite{Breiman:2001} are two measures of variable importance in random forest based on out-of-bag assessment and node impurities, respectively.

The aforementioned  SOIL can be generalized to  models with categorical variables. For each element $M$ in $\mathcal{M}$, we calculate the corresponding weight $w_M$ below according to the BIC-p weighting \citep{Nan:Yang:2014}, which is defined as
\beqr\label{BIC_weight}
w_M\defby\exp{\Big(-\frac{I_M}{2}-\psi C_M\Big)}/\sum\limits_{M'\in \mathcal{M}}\exp{\Big(-\frac{I_{M'}}{2}-\psi C_{M'}\Big)}, \; \mbox{for}\;M\in \mathcal{M},
\eeqr
where $I_M$ is the BIC information criterion for model $M$, $C_M=|M|\log{({e\cdot p^{\ast}}/{|M|})}+2\log{(|M|+2)}$, $ |\cdot|$ denotes cardinality, and $\psi$ is a positive constant. According to the model set $\mathcal{M}=\{M_i: i=1,\ldots,N\}$ and weighting vector $w=\left(w_{1}, \dots, w_{N}\right)^{\mathrm{T}}$ defined in \eqref{BIC_weight},  the importance of the $j$th variable $X_j$, $j\in\{1,\ldots,p\}$, is defined as
\begin{equation}\label{defi SOIL}
S_{j}\stackrel{\mbox{\rm{\tiny def}}}{=} S(j ; w, \mathcal{M})=\sum_{i=1}^{N} w_{i} I\left(\mathcal{I}_j \subseteq {M}_{i}\right),
\end{equation}
where $I(\cdot)$ denotes the indicator function.
Thus, the importance of the $j$th variable $X_j$ is the sum of weights of the candidate models including the variable $X_j$. The choice of  $\psi$ can be specified by the users. Choosing  too small a $\psi$ may result in significant SOIL importances of unimportant variables and vice versa. As suggested in \cite{Ye:Yi:Yang:2018}, SOIL under $\psi=0.5$ or $\psi=1$ generally performs well. In practice, it is computationally impractical to consider all subsets as candidate model set $\mathcal{M}$ for $p\gg n$.
Our implementation strategy is to make use of three group variable selection methods (e.g., group LASSO \citep{Yuan:Lin:2006}, group SCAD \citep{Wang:Chen:Li:2007}, and group MCP \citep{Huang:Breheny:Ma:2012}) first, and put  together the resultant models with various levels of sparsity to obtain the candidate models.

SOIL has a desirable theoretical property, that is, it can well separate the true predictors from the others asymptotically \citep{Ye:Yi:Yang:2018} under the condition of  consistency of the weighting. \cite{Chen:Zhang:Xu:Yang:2021} has demonstrated the rationality of this consistency condition based on BIC-p weighting. Thus, SOIL is a recommendable method in both theory and practical application.

In this work, with SOIL, we propose a variable importance based interaction modeling (VIBIM) procedure for linear models with categorical predictors. The procedure consists of the following four steps:

\textit {Step} 1. Using \eqref{defi SOIL} to compute the SOIL importances  $S_1,\ldots,S_p$ based on the linear candidate models without  any interaction term.

\textit {Step} 2. Set up a threshold value $c \in (0,1)$, and define a  submodel
$
M_{c}=\{i:S_i > c, \mbox{for} \,1\leqslant i \leqslant p\}.
$
Then add the  interactions $\{X_{\mathcal{I}_{i}}X_{\mathcal{I}_{j}}:i,j \in M_{c}\}$ to $\{X_{\mathcal{I}_{1}},\ldots,X_{\mathcal{I}_{p}}\}$, where $X_{\mathcal{I}_{i}}X_{\mathcal{I}_{j}}$ denotes  the vector consisting of all pairwise products of elements of $X_{\mathcal{I}_{i}}$ and $X_{\mathcal{I}_{j}}$
 with length $|\mathcal{I}_{i}| \cdot |\mathcal{I}_{j}| $. For example, $X_{\mathcal{I}_{i}}X_{\mathcal{I}_{j}}=(X_{i,1}X_{j,1},\ldots, X_{i,1}X_{j,J_j-1},\ldots, $ \\$ X_{i,J_i-1}X_{j,1},\ldots,X_{i,J_i-1}X_{j,J_j-1})^{\mathrm{T}}$ for two categorical variables.

\textit {Step} 3. Recompute the SOIL importance for the augmented dataset to sort the $p$ main effects together with $\frac{1}{2}|M_c| \cdot (|M_c|-1)$ interactions $\{X_{\mathcal{I}_{i}}X_{\mathcal{I}_{j}}:i,j \in M_{c}, i \neq j\}$  in a descending order $S_{j_1}^{\prime} \geq\cdots \geq S_{j_{p^{\prime}}}^{\prime}$, where $p^{\prime}=p+\frac{1}{2}|M_c| \cdot (|M_c|-1)$.

\textit {Step} 4. Obtain a sequence of nested models $M_{(1)}\subseteq \cdots \subseteq M_{(K)}$,  the $s$th one of
which includes the first $s$ predictors according to the SOIL importance calculated in \textit {Step} 3.

\textit {Step} 5. Narrow down the above nested models. One can choose a small model $M_{(L)}$  by minimizing BIC and a large model $M_{(U)}$ by minimizing AIC, and the models  between $M_{(L)}$ and $M_{(U)}$ are considered as most plausible. In high-dimensional situations, we can use  AIC-p (see, e.g., \cite{yang:1999:AIC}) and BIC-p information criteria to avoid overfitting.

Unlike  the existing model selection methods that rely on a single final model, and  model confidence set approaches that typically give out too many models, VIBIM delivers a relatively small number of nested models. We point out that the most appropriate threshold value $c $  and the number of nested models $K$ are not specified,  since they  may depend on the goal of the analysis and application. Model selection methods may suffer from  the disadvantage of  excluding predictors that are highly correlated with ones that are already in a model, whether or not they should be included. Our method alleviates this problem, as we will see in Section 5, where we identify a  significant interaction that is highly correlated with its main effect. By taking advantage of model averaging and variable importance, VIBIM can yield good models in terms of interpretability,  stability, reliability and prediction.
\section{Simulation Study}
We demonstrate the  performance of interaction selection of the VIBIM method in various high-dimensional scenarios. For all the numerical  analysis in this paper, we  obtain the candidate models using the R package \texttt{grpreg} with the default settings, which determines a grid of 100  tuning parameters that are equally spaced on the log scale  over a range $[\lambda_{\min}, \lambda_{\max}]$  (see, e.g., \cite{Breheny:Huang:2011} and \cite{Huang:Breheny:Ma:2012} for choices of $\lambda_{\min}$ and $\lambda_{\max}$). We choose $\psi=1$ in (2) for the BIC-p weighting and  the threshold value $c=$ 1e-4 in \textit {Step} 2 for VIBIM procedure. For comparison, we also consider three  procedures: group LASSO, group SCAD, and group MCP (abbreviated as gLASSO, gSCAD, gMCP, respectively). The gLASSO first selects main effects  by 10-fold cross-validation, then identifies main effects and important interactions among the reduced list of main effects. The gSCAD and gMCP are conducted similarly.

We generate the simulation data by the linear model (1) with simple size $n$, where the error $\epsilon \sim N(0,\sigma^2)$. Let $(Z_1,\ldots,Z_p)\sim N_p(0,\Sigma)$ with $\Sigma=(\rho^{|i-j|})_{p\times p}$. Notice that $X_{\mathcal{I}_{i}}=(X_{i,1},\ldots,X_{i,J_i-1})^{\mathrm{T}}$ for $i=1,\ldots,q$, let $X_{i,j}=I(\Phi_{(j-1)/J_i}<Z_i\leq\Phi_{{j/J_i}})$ for $j=1,\ldots,J_i-1$, where $\Phi_{(j-1)/J_i}$ is the $(j-1)/J_i$-th quantile of the standard normal distribution. Let $X_i=Z_i$ for $i\in\{q+1,\ldots,p\}$. In all the examples, we set $\beta_0=1$, $(\rho,\sigma)=(0.5,1)$. The number of categorical variables is $q=6$ and $J_1=J_2=J_3=J_4=2$, $J_5=J_6=6$. We choose the threshold value $c=$ 1e-4 in \textit {Step} 2 of VIBIM procedure.
Three different settings of $(n,p)$ are considered: $(100, 1000)$, $(200, 1000)$, $(3000, 1000)$, the results of $(n,p)=(100, 1000)$ and $(n,p)=(300, 1000)$ are  provided in the supplementary materials due to the space limitation. All simulation examples are repeated  100 times and the corresponding values are averaged.

Two examples are considered in the simulations. First, we consider  interaction models that obey strong heredity, i.e., the interactions are only among pairs of nonzero main effects. Second, we simulate interaction models that obey weak heredity, i.e., each interaction has only one of its main effects present.

\noindent\textbf{Example 1 }(Strong heredity). We first consider the following six interaction models:

Model \Rmnum{1}: $Y=\beta_{\mathcal{I}_1}^{\mathrm{T}}X_{\mathcal{I}_1}+\beta_{\mathcal{I}_3}^{\mathrm{T}}X_{\mathcal{I}_3}
+\beta_{\mathcal{I}_5}^{\mathrm{T}}X_{\mathcal{I}_5}+X_7 \beta_7+X_8 \beta_8+X_9 \beta_9+\beta_{7,9}X_7X_9+\epsilon$,

 Model \Rmnum{2}: $Y=\beta_{\mathcal{I}_1}^{\mathrm{T}}X_{\mathcal{I}_1}+\beta_{\mathcal{I}_3}^{\mathrm{T}}X_{\mathcal{I}_3}
+\beta_{\mathcal{I}_5}^{\mathrm{T}}X_{\mathcal{I}_5}+X_7 \beta_7+X_8 \beta_8+X_9 \beta_9+\beta_{1,8}^{\mathrm{T}}X_{\mathcal{I}_1}X_8+\epsilon$,

Model \Rmnum{3}: $Y=\beta_{\mathcal{I}_1}^{\mathrm{T}}X_{\mathcal{I}_1}+\beta_{\mathcal{I}_3}^{\mathrm{T}}X_{\mathcal{I}_3}
+\beta_{\mathcal{I}_5}^{\mathrm{T}}X_{\mathcal{I}_5}+X_7 \beta_7+X_8 \beta_8+X_9 \beta_9+\beta_{1,3}^{\mathrm{T}}X_{\mathcal{I}_1}X_{\mathcal{I}_3}+\epsilon.$

Model \Rmnum{4}:
$Y=\beta_{\mathcal{I}_1}^{\mathrm{T}}X_{\mathcal{I}_1}+\beta_{\mathcal{I}_3}^{\mathrm{T}}X_{\mathcal{I}_3}
+\beta_{\mathcal{I}_5}^{\mathrm{T}}X_{\mathcal{I}_5}+X_7 \beta_7+X_8 \beta_8+X_9 \beta_9+\beta_{7,9}X_7X_9+\beta_{1,8}^{\mathrm{T}}X_{\mathcal{I}_1}X_8+\epsilon.$

Model \Rmnum{5}:
$Y=\beta_{\mathcal{I}_1}^{\mathrm{T}}X_{\mathcal{I}_1}+\beta_{\mathcal{I}_3}^{\mathrm{T}}X_{\mathcal{I}_3}
+\beta_{\mathcal{I}_5}^{\mathrm{T}}X_{\mathcal{I}_5}+X_7 \beta_7+X_8 \beta_8+X_9 \beta_9+\beta_{7,9}X_7X_9+\beta_{1,3}^{\mathrm{T}}X_{\mathcal{I}_1}X_3+\epsilon.$

Model \Rmnum{6}:
$Y=\beta_{\mathcal{I}_1}^{\mathrm{T}}X_{\mathcal{I}_1}+\beta_{\mathcal{I}_3}^{\mathrm{T}}X_{\mathcal{I}_3}
+\beta_{\mathcal{I}_5}^{\mathrm{T}}X_{\mathcal{I}_5}+X_7 \beta_7+X_8 \beta_8+X_9 \beta_9+\beta_{7,9}X_7X_9+\beta_{1,8}^{\mathrm{T}}X_{\mathcal{I}_1}X_8+\beta_{1,3}^{\mathrm{T}}X_{\mathcal{I}_1}X_{\mathcal{I}_3}+\epsilon.$

\noindent The true $(\beta_{1},\beta_{3},\beta_{7},\beta_{8},\beta_{9})=(2,3,2,3,-2)$, $\beta_{\mathcal{I}_5}=(-2,-3,-4,-5,0)^{\mathrm{T}}$.  The coefficients of the interactions are $\beta_{7,9}=1.5$, $\beta_{1,8}=1.5$, $\beta_{1,3}=2$. The first three models Model \Rmnum{1}$-$\Rmnum{3} include an interaction between two continuous variables, a continuous variable and a categorical variable and two categorical variables, respectively. Models \Rmnum{4} and \Rmnum{5} include two interactions and Model \Rmnum{6} contains three interactions.

\noindent\textbf{Example 2} (Weak heredity).

Model \Rmnum{1}: $Y=\beta_{\mathcal{I}_1}^{\mathrm{T}}X_{\mathcal{I}_1}+\beta_{\mathcal{I}_3}^{\mathrm{T}}X_{\mathcal{I}_3}
+\beta_{\mathcal{I}_5}^{\mathrm{T}}X_{\mathcal{I}_5}+X_7 \beta_7+X_8 \beta_8+\beta_{7,9}X_7X_9+\epsilon$,

 Model \Rmnum{2}: $Y=\beta_{\mathcal{I}_1}^{\mathrm{T}}X_{\mathcal{I}_1}+\beta_{\mathcal{I}_3}^{\mathrm{T}}X_{\mathcal{I}_3}+\beta_{\mathcal{I}_5}^{\mathrm{T}}X_{\mathcal{I}_5}
+X_7 \beta_7+X_9 \beta_9+\beta_{1,8}^{\mathrm{T}}X_{\mathcal{I}_1}X_8+\epsilon$.

Model \Rmnum{3}: $Y=\beta_{\mathcal{I}_3}^{\mathrm{T}}X_{\mathcal{I}_3}+\beta_{\mathcal{I}_5}^{\mathrm{T}}X_{\mathcal{I}_5}
+X_7 \beta_7+X_8 \beta_8+X_9 \beta_9+\beta_{1,8}^{\mathrm{T}}X_{\mathcal{I}_1}X_8+\epsilon$.

Model \Rmnum{4}: $Y=\beta_{\mathcal{I}_1}^{\mathrm{T}}X_{\mathcal{I}_1}+\beta_{\mathcal{I}_5}^{\mathrm{T}}X_{\mathcal{I}_5}+X_7 \beta_7+X_8 \beta_8+X_9 \beta_9+\beta_{1,3}^{\mathrm{T}}X_{\mathcal{I}_1}X_{\mathcal{I}_3}+\epsilon$.

\noindent We adopt the same  setup of  coefficients as in Example 1. In Model \Rmnum{1}, we consider an interaction among two continuous variables with only one  parent effect.
Model \Rmnum{2} contains  an interaction between  a continuous variable and a categorical variable  which has only its categorical main effect present, whereas  Model \Rmnum{3} includes this interaction and its continuous main effect. In the  last model,   an interaction between two categorical variables with only one  parent effect are considered.

To evaluate the variable selection performance of each method, we employ two performance measures, $F$- and $G$-measures. The $F$-measure for a given model $M_{0}$ is defined as the harmonic mean of the precision and recall, and the $G$-measure is defined as the geometric mean of the two. Since we focus on the task of interaction selection, the precision and recall are calculated with respect to interaction terms.  Specifically,
$
F\left(M_{0}\right)=2|\widetilde{M}_{0} \cap M^{*}|/(|\widetilde{M}_{0}|+|M^{*}|)
$
and
$
G\left(M_{0}\right)=|\widetilde{M}_{0} \cap M^{*}|/\sqrt{|\widetilde{M}_{0}| \cdot\left|M^{*}\right|},
$
where $\widetilde{M}_0$ denotes interaction terms in $M_0$ and $M^{*}$ represents interactions in the true model. To ensure a fair comparison, we present simulation results for multiple models $\rm {M_{(i)}},i=5,\ldots,10$ consisting of $i$ predictors obtained by VIBIM, gLASSO, gSCAD, gMCP, respectively. To be specific, for VIBIM, $\rm {M_{(i)}},i=5,\ldots,10$ stand for the nested models in $Step$ 4, and for  three variable selection methods, $\rm {M_{(i)}},i=5,\ldots,10$ are obtained from solution path which have the model size $i=5,\ldots,10$, respectively.

The means and standard errors (in parentheses) of $F$- and $G$-measures for Examples 1$-$2 are summarized in Tables \ref{example1}$-$\ref{example2_1}. It is seen that when the strong heredity assumption holds, the VIBIM performs best, across all of the settings, with the highest  $F$- and $G$-measures. The $F$- and $G$-measures are generally very close to 1 when the model size of VIBIM  is around the true model size. In Example 2, when the weak heredity condition holds, the VIBIM outperforms the other methods in Model \Rmnum{2} and Model \Rmnum{4}. In  Model \Rmnum{1} and Model \Rmnum{3}, however, no method could identify correctly the important interactions. This is perhaps because the low correlation between the interaction and its parent effects. For instance,  the Pearson correlation coefficient between $X_{\mathcal{I}_1}X_8$ and $X_{8}$ is around 0.709, and thus $X_{8}$ is typically identified as  an important main effect in the first stage, while that between $X_{\mathcal{I}_1}X_8$ and $X_{\mathcal{I}_1}$ is only around $-0.003$. Therefore, in Example 2, the  four methods tend to select $X_{\mathcal{I}_1}X_8$ in Model \Rmnum{2}, but fail in Model \Rmnum{3}. The gMCP  performs mostly well from the perspective of $G$-measure, but less so in  $F$-measure.  Overall, the VIBIM is the best among all the methods in terms of $F$- and $G$-measures.

We also conduct another version of  the procedures based on the three variable selection methods, i.e., the selected main effects are obtained from  the solution  path of each of the selection methods at the first stage,  and the number of  the main effects  are all equal to the cardinality of $M_c$ in $Step$ 2 of VIBIM. These results  can be found in Section S.1 of the Supplementary Material, which shows that our proposed  method has a  clear advantage over the other methods. In addition,
in Supplementary Material Section  S.1, we present  simulation results of  importance measures SOIL, RFI1 and RFI2 for Examples 1$-$2 at the first stage of VIBIM procedure.

\begin{table}
\setlength{\belowcaptionskip}{0.3cm}
\caption{\label{example1} Results for the Models \Rmnum{1}$-$\Rmnum{6} that are considered  in Example 1.} \vspace{0cm}
\centering
\linespread{0.95}
\centering
\resizebox{0.9\textwidth}{!}{\footnotesize{
\begin{tabular}{llHccccHcccccc}
\toprule
&Method   & \multicolumn{5}{c}{$F$-measure}            & \multicolumn{5}{c}{$G$-measure}  \\
\cmidrule(r){3-7}   \cmidrule(l){8-12}
& &$\mathrm{M}_{(6)}$  &$\mathrm{M}_{(7)}$ &$\mathrm{M}_{(8)}$ &$\mathrm{M}_{(9)}$
&$\mathrm{M}_{(10)}$  &$\mathrm{M}_{(6)}$ &$\mathrm{M}_{(7)}$ &$\mathrm{M}_{(8)}$ &$\mathrm{M}_{(9)}$ &$\mathrm{M}_{(10)}$ \\\hline
\multicolumn{12}{c}{Model \Rmnum{1}}\\
&$\mathrm{VIBIM}$  &0.240   & 1.000   & 0.967   & 0.932   & 0.840   & 1.000   & 1.000   & 0.971   & 0.940   & 0.860  \\
&& (0.043) & (0.000)   & (0.010) & (0.014) & (0.017) & (0.000)   & (0.000)   & (0.009) & (0.012) & (0.015)\\
&$\mathrm{gLASSO}$ &0.556   & 0.552   & 0.546   & 0.441   & 0.402   & 0.702   & 0.677   & 0.623   & 0.576   & 0.532 \\
& &(0.031) & (0.027) & (0.017) & (0.019) & (0.015) & (0.019) & (0.017) & (0.013) & (0.009) & (0.007) \\
&$\mathrm{gSCAD}$  & 0.627   & 0.703   & 0.668   & 0.554   & 0.464   & 0.711   & 0.740   & 0.718   & 0.743   & 0.705 \\
& &(0.027) & (0.020) & (0.017) & (0.034) & (0.035) & (0.020) & (0.018) & (0.013) & (0.017) & (0.018) \\
&$\mathrm{gMCP}$ &0.814   & 0.873   & 0.823   & 0.693   & 0.617   & 0.851   & 0.889   & 0.972   & 0.945   & 0.943   \\
& &(0.021) & (0.016) & (0.036) & (0.043) & (0.046) & (0.015) & (0.014) & (0.009) & (0.012) & (0.012) \\ \hline
\multicolumn{12}{c}{Model \Rmnum{2}}\\
&$\mathrm{VIBIM}$   &  0.190   & 1.000   & 0.920   & 0.840   & 0.755   & 1.000   & 1.000   & 0.930   & 0.860   & 0.788 \\
&  &  (0.039) & (0.000)   & (0.014) & (0.018) & (0.019) & (0.000)   & (0.000)   & (0.013) & (0.015) & (0.016)    \\
&$\mathrm{gLASSO}$ &  0.565   & 0.583   & 0.577   & 0.478   & 0.388   & 0.676   & 0.679   & 0.660   & 0.587   & 0.548 \\
& & (0.028) & (0.026) & (0.020) & (0.017) & (0.020) & (0.021) & (0.018) & (0.013) & (0.010) & (0.009)   \\
&$\mathrm{gSCAD}$  & 0.178   & 0.265   & 0.597   & 0.557   & 0.408   & 0.209   & 0.272   & 0.676   & 0.754   & 0.743      \\
& &  (0.030) & (0.042) & (0.039) & (0.037) & (0.039) & (0.033) & (0.042) & (0.035) & (0.022) & (0.020) \\
&$\mathrm{gMCP}$ & 0.117   & 0.887   & 0.747   & 0.620   & 0.485   & 0.280   & 0.898   & 0.940   & 0.935   & 0.898  \\
& & (0.026) & (0.020) & (0.040) & (0.045) & (0.046) & (0.036) & (0.018) & (0.015) & (0.013) & (0.016)\\ \hline
&   &$\mathrm{M}_{(7)}$ &$\mathrm{M}_{(8)}$ &$\mathrm{M}_{(9)}$
&$\mathrm{M}_{(10)}$  &$\mathrm{M}_{(11)}$ &$\mathrm{M}_{(7)}$ &$\mathrm{M}_{(8)}$ &$\mathrm{M}_{(9)}$ &$\mathrm{M}_{(10)}$ &$\mathrm{M}_{(11)}$ \\\hline
\multicolumn{12}{c}{Model \Rmnum{3}}\\
&$\mathrm{VIBIM}$  &1.000   & 0.917   & 0.852   & 0.800   & 0.761   & 1.000   & 0.927   & 0.870   & 0.826   & 0.793   \\
&& (0.000)   & (0.015) & (0.018) & (0.019) & (0.020) & (0.000)   & (0.013) & (0.016) & (0.016) & (0.017) \\
&$\mathrm{gLASSO}$  & 0.857   & 0.706   & 0.538   & 0.462   & 0.357   & 0.898   & 0.762   & 0.655   & 0.593   & 0.528   \\
& &(0.022) & (0.020) & (0.021) & (0.020) & (0.020) & (0.014) & (0.013) & (0.009) & (0.009) & (0.009) \\
&$\mathrm{gSCAD}$ &0.920   & 0.643   & 0.463   & 0.415   & 0.440   & 0.971   & 0.840   & 0.747   & 0.632   & 0.687   \\
& &(0.025) & (0.046) & (0.046) & (0.043) & (0.042) & (0.012) & (0.034) & (0.037) & (0.039) & (0.033) \\
&$\mathrm{gMCP}$ &0.720   & 0.530   & 0.523   & 0.505   & 0.507   & 0.960   & 0.885   & 0.866   & 0.880   & 0.900   \\
& & (0.045) & (0.049) & (0.048) & (0.047) & (0.048) & (0.020) & (0.030) & (0.029) & (0.024) & (0.022) \\ \hline
\multicolumn{12}{c}{Model \Rmnum{4}}\\
&$\mathrm{VIBIM}$   &  0.727   & 1.000   & 0.966   & 0.930   & 0.879   & 0.760   & 1.000   & 0.969   & 0.936   & 0.891   \\
&  &  (0.013) & (0.000)   & (0.008) & (0.010) & (0.012) & (0.011) & (0.000)   & (0.007) & (0.009) & (0.011) \\
&$\mathrm{gLASSO}$ &  0.654   & 0.645   & 0.646   & 0.605   & 0.494   & 0.756   & 0.759   & 0.734   & 0.698   & 0.656   \\
& & (0.027) & (0.028) & (0.022) & (0.021) & (0.025) & (0.014) & (0.014) & (0.011) & (0.010) & (0.008) \\
&$\mathrm{gSCAD}$  & 0.560   & 0.596   & 0.676   & 0.610   & 0.470   & 0.573   & 0.605   & 0.740   & 0.776   & 0.764   \\
& & (0.017) & (0.023) & (0.030) & (0.033) & (0.038) & (0.017) & (0.023) & (0.024) & (0.015) & (0.014) \\
&$\mathrm{gMCP}$ & 0.647   & 0.933   & 0.700   & 0.614   & 0.421   & 0.673   & 0.937   & 0.950   & 0.949   & 0.923   \\
& &(0.011) & (0.012) & (0.042) & (0.046) & (0.046) & (0.012) & (0.011) & (0.008) & (0.008) & (0.010)  \\ \hline
&   &$\mathrm{M}_{(8)}$ &$\mathrm{M}_{(9)}$
&$\mathrm{M}_{(10)}$  &$\mathrm{M}_{(11)}$ &$\mathrm{M}_{(12)}$ &$\mathrm{M}_{(8)}$ &$\mathrm{M}_{(9)}$ &$\mathrm{M}_{(10)}$ &$\mathrm{M}_{(11)}$ &$\mathrm{M}_{(12)}$ \\\hline
\multicolumn{12}{c}{Model \Rmnum{5}}\\
&$\mathrm{VIBIM}$  &  1.000   & 0.974   & 0.955   & 0.927   & 0.901   & 1.000   & 0.976   & 0.959   & 0.933   & 0.910   \\
&&(0.000)   & (0.007) & (0.009) & (0.010) & (0.011) & (0.000)   & (0.006) & (0.008) & (0.009) & (0.010) \\
&$\mathrm{gLASSO}$ &0.780   & 0.743   & 0.581   & 0.537   & 0.438   & 0.870   & 0.811   & 0.743   & 0.675   & 0.630   \\
& & (0.028) & (0.022) & (0.029) & (0.023) & (0.025) & (0.013) & (0.010) & (0.008) & (0.007) & (0.005) \\
&$\mathrm{gSCAD}$  &0.886   & 0.725   & 0.600   & 0.549   & 0.422   & 0.938   & 0.824   & 0.766   & 0.742   & 0.707   \\
& &(0.024) & (0.032) & (0.035) & (0.036) & (0.037) & (0.011) & (0.020) & (0.020) & (0.020) & (0.020) \\
&$\mathrm{gMCP}$ & 0.878   & 0.717   & 0.706   & 0.585   & 0.539   & 0.925   & 0.864   & 0.862   & 0.884   & 0.867   \\
& &   (0.023) & (0.036) & (0.036) & (0.043) & (0.044) & (0.013) & (0.018) & (0.018) & (0.015) & (0.017) \\ \hline
\multicolumn{12}{c}{Model \Rmnum{6}}\\
&$\mathrm{VIBIM}$   &  0.856   & 0.998   & 0.977   & 0.960   & 0.947   & 0.868   & 0.998   & 0.978   & 0.962   & 0.951   \\
&  &  (0.009) & (0.002) & (0.005) & (0.007) & (0.008) & (0.008) & (0.002) & (0.005) & (0.007) & (0.008) \\
&$\mathrm{gLASSO}$ &  0.826   & 0.809   & 0.772   & 0.692   & 0.616   & 0.903   & 0.888   & 0.843   & 0.792   & 0.741   \\
& &(0.027) & (0.027) & (0.024) & (0.025) & (0.025) & (0.012) & (0.012) & (0.010) & (0.008) & (0.006) \\
&$\mathrm{gSCAD}$  &0.796   & 0.813   & 0.760   & 0.560   & 0.554   & 0.817   & 0.835   & 0.815   & 0.811   & 0.775   \\
& & (0.020) & (0.020) & (0.024) & (0.039) & (0.036) & (0.016) & (0.016) & (0.015) & (0.014) & (0.014) \\
&$\mathrm{gMCP}$ & 0.953   & 0.896   & 0.694   & 0.571   & 0.474   & 0.965   & 0.947   & 0.907   & 0.898   & 0.901   \\
& & (0.013) & (0.023) & (0.039) & (0.044) & (0.046) & (0.008) & (0.010) & (0.011) & (0.012) & (0.012)  \\
\bottomrule
\end{tabular}}}
\vspace{-0.5cm}
\end{table}

\begin{table}
\setlength{\belowcaptionskip}{0.3cm}
\caption{\label{example2_1} Results for the Models \Rmnum{1}$-$\Rmnum{4} that are considered  in Example 2.} \vspace{0cm}
\linespread{1}
 \centering
\resizebox{0.9\textwidth}{!}{\footnotesize{
\begin{tabular}{llHccccHcccccc}
\toprule
&Method   & \multicolumn{5}{c}{$F$-measure}            & \multicolumn{5}{c}{$G$-measure}  \\
\cmidrule(r){3-7}   \cmidrule(l){8-12}
& &$\mathrm{M}_{(5)}$ &$\mathrm{M}_{(6)}$ &$\mathrm{M}_{(7)}$ &$\mathrm{M}_{(8)}$ &$\mathrm{M}_{(9)}$
&$\mathrm{M}_{(5)}$ &$\mathrm{M}_{(6)}$ &$\mathrm{M}_{(7)}$ &$\mathrm{M}_{(8)}$ &$\mathrm{M}_{(9)}$\\\hline
\multicolumn{12}{c}{Model \Rmnum{1}}\\
&$\mathrm{VIBIM}$   &0.000   & 0.030   & 0.030   & 0.030   & 0.027   & 0.000   & 0.031   & 0.030   & 0.030   & 0.027 \\
&  & (0.000)   & (0.017) & (0.017) & (0.017) & (0.015) & 0.000   & (0.017) & (0.017) & (0.017) & (0.016)  \\
&$\mathrm{gLASSO}$ &0.103   & 0.100   & 0.082   & 0.047   & 0.047   & 0.124   & 0.111   & 0.098   & 0.061   & 0.063 \\
& &(0.028) & (0.027) & (0.023) & (0.015) & (0.014) & (0.030) & (0.028) & (0.025) & (0.018) & (0.018)   \\
&$\mathrm{gSCAD}$  &0.030   & 0.033   & 0.027   & 0.015   & 0.005   & 0.037   & 0.036   & 0.032   & 0.018   & 0.008    \\
& &  (0.015) & (0.017) & (0.013) & (0.011) & (0.005) & (0.017) & (0.018) & (0.015) & (0.012) & (0.007) \\
&$\mathrm{gMCP}$ & 0.010   & 0.010   & 0.010   & 0.000   & 0.010   & 0.015   & 0.010   & 0.011   & 0.000   & 0.013  \\
& & (0.010) & (0.010) & (0.010) & (0.000)   & (0.010) & (0.012) & (0.010) & (0.011) & (0.000)    & (0.011) \\ \hline
\multicolumn{12}{c}{Model \Rmnum{2}}\\
&$\mathrm{VIBIM}$   &  0.200   & 0.980   & 0.913   & 0.847   & 0.774   & 1.000   & 0.980   & 0.921   & 0.864   & 0.802  \\
&  &  (0.040) & (0.014) & (0.019) & (0.021) & (0.023) & 0.000   & (0.014) & (0.018) & (0.020) & (0.020)    \\
&$\mathrm{gLASSO}$ &  0.125   & 0.233   & 0.355   & 0.420   & 0.342   & 0.139   & 0.280   & 0.396   & 0.505   & 0.489  \\
& & (0.029) & (0.033) & (0.031) & (0.022) & (0.019) & (0.031) & (0.036) & (0.033) & (0.022) & (0.014)  \\
&$\mathrm{gSCAD}$  & 0.107   & 0.407   & 0.665   & 0.647   & 0.470   & 0.128   & 0.425   & 0.722   & 0.767   & 0.725    \\
& & (0.026) & (0.041) & (0.022) & (0.032) & (0.037) & (0.029) & (0.042) & (0.018) & (0.020) & (0.020)  \\
&$\mathrm{gMCP}$ & 0.450   & 0.830   & 0.697   & 0.692   & 0.585   & 0.704   & 0.854   & 0.903   & 0.878   & 0.833    \\
& & (0.041) & (0.023) & (0.041) & (0.040) & (0.043) & (0.031) & (0.021) & (0.019) & (0.020) & (0.024)  \\ \hline
\multicolumn{12}{c}{Model \Rmnum{3}}\\
&$\mathrm{VIBIM}$  &0.000   & 0.000   & 0.000   & 0.000   & 0.000   & 0.0000      & 0.000   & 0.000   & 0.000   & 0.000
  \\
&  & (0.000)   & (0.000)   & (0.000)   & (0.000)   & (0.000)   & (0.000)   & (0.000)   & (0.000)   & (0.000)   & (0.000) \\
&$\mathrm{gLASSO}$ &0.040   & 0.043   & 0.038   & 0.037   & 0.022   & 0.190   & 0.068   & 0.053   & 0.047   & 0.033  \\
& &(0.020) & (0.019) & (0.017) & (0.017) & (0.011) & (0.040) & (0.024) & (0.020) & (0.019) & (0.014) \\
&$\mathrm{gSCAD}$  & 0.000   & 0.010   & 0.000   & 0.007   & 0.007   & 0.000   & 0.032   & 0.000   & 0.014   & 0.014 \\
& &(0.000)   & (0.010) & (0.000)   & (0.007) & (0.007) & (0.000)   & (0.018) & (0.000)   & (0.010) & (0.010) \\
&$\mathrm{gMCP}$ &0.000   & 0.000   & 0.000   & 0.000   & 0.000   & 0.000      & 0.000   & 0.000   & 0.000   & 0.000      \\
&  & (0.000)   & (0.000)   & (0.000)   & (0.000)   & (0.000)   & (0.000)   & (0.000)   & (0.000)   & (0.000)   & (0.000) \\ \hline
\multicolumn{12}{c}{Model \Rmnum{4}}\\
&$\mathrm{VIBIM}$   &  0.160   & 1.000   & 0.923   & 0.863   & 0.811   & 1.000   & 1.000   & 0.933   & 0.881   & 0.836  \\
&  &  (0.037) & 0.000   & (0.014) & (0.018) & (0.020) & 0.000   & 0.000   & (0.012) & (0.016) & (0.017)    \\
&$\mathrm{gLASSO}$ &  0.727   & 0.823   & 0.658   & 0.515   & 0.430   & 0.915   & 0.884   & 0.743   & 0.651   & 0.588  \\
& & (0.041) & (0.032) & (0.023) & (0.024) & (0.023) & (0.021) & (0.024) & (0.014) & (0.011) & (0.011)   \\
&$\mathrm{gSCAD}$  &0.837   & 0.923   & 0.757   & 0.625   & 0.518   & 0.976   & 0.947   & 0.866   & 0.802   & 0.768 \\
& & (0.035) & (0.022) & (0.033) & (0.035) & (0.038) & (0.008) & (0.017) & (0.019) & (0.016) & (0.019)\\
&$\mathrm{gMCP}$ & 0.360   & 0.940   & 0.770   & 0.752   & 0.633   & 0.973   & 1.000   & 0.946   & 0.927   & 0.895  \\
& & (0.048) & (0.024) & (0.038) & (0.038) & (0.042) & (0.016) & 0.000   & (0.011) & (0.013) & (0.015) \\
\bottomrule
\end{tabular}}}
\vspace{-0.5cm}
\end{table}
\section{COVID-19 Data and Methods}
\label{sec:data}
\subsection{Data Source and Modification}
\label{sec:Processing}
Based on a dataset about COVID-19 with 296 cities across China, \cite{Tian:2020} explored the associations between the number of confirmed cases  (\textit{Sevendays.Cucase})  during the first week of the outbreak (i.e. reported the first case) in a city and three transmission control measures: closing entertainment venues, banning intra-city public transport and prohibition of railway from and to other cities. For the sake of description, we denote the three control measures as \textit{Enter}, \textit{Bus} and \textit{Railway}, respectively.
Specifically, using Poisson regression model, \cite{Tian:2020} investigated the associations between \textit{Sevendays.Cucase} and 10 explanatory variables: a binary variable indicating whether  control measure  $K$ $ (K=\textit{Enter}, \textit{Bus}, \textit{Railway})$ was implemented in city $i$ before or during its first week of outbreak ($\textit{K.Resp}^*$), the timing of implementing transmission control measure $K$ in the city, the date of the first confirmed case in the city (\textit{Arr.Time}), the distance from Wuhan City (\textit{Dis.WH}), the population of the city in 2018 (\textit{Pop.2018}), and the passenger volume from Wuhan in 2018 (\textit{Total.Flow}). A more detailed description of these variables can be found in Table \ref{t1}.

Upon a careful examination, we feel there may be several issues in their analysis. First, the causal relationships between variables $\textit{K.Resp}^*$ and \textit{Sevendays.Cucase}, and between variables $\textit{K.Date}^*$ and \textit{Sevendays.Cucase} are not clear. More precisely, the percent of cities that implemented at least one measure after the outbreak is $208/296=70.27\%$. Such measures may have been implemented as a response to, rather than a prevention of, the arrival of COVID-19 in most cities. In other words, the causality confusion arises because
of a possible opposite causal connection: the response variable may be a causal factor in explanatory variables related to control measures. Second, the  definition of  $K.Date^{*}$ is questionable. Recall the definition of $K.Date_i^{*}$, which is the  timing of implementing control measure K in city $i$, where 31
December 2019 is day 0, if $K.Resp_i^{*}=1$; $K.Date_i^{*}=0$ if $K.Resp_i^{*}=0$.
 We argue that the setting that  $K.Date_i^{*}=0$ when $K.Resp_i^{*}=0$ has the potential to mislead. Since $K.Date_i^{*}=0$ suggests that city $i$ did not implement $K$ in practice before or during its first week of outbreak, but measure $K$ may be implemented after  first week of outbreak in city $i$, and in this case, $K.Date^{*}_i$ is in fact greater than 0 by the definition of $K.Date_i^{*}$, which seems contradictory.
 Third, quite surprisingly, \cite{Tian:2020} drew a conclusion that $Dis.WH$ had a significant
positive effect on the response variable, which indicates that cities farther away from Wuhan had more confirmed cases when the other covariates were controlled.

\vspace{-0.1cm}
In view of these issues, we modify variable $\textit{K.Resp}^*$ as the variable $\textit{K.Resp}$ indicating whether measure $K$ has been implemented before the first case was reported. In the meanwhile, variable \textit{K.Date} represents the number of days that measure $K$ was implemented before the outbreak of COVID-19.  In addition, the puzzling sign issue of $Dis.WH$ might be due to improper modeling,  or might be caused by omitting some important variables   in light of the weak
 correlations between $Dis.WH$ and the other explanatory variables. Therefore, we supplement 4 continuous explanatory variables: average travel intensity (\textit{Travel.Intensity}), per capita gross regional product (\textit{PGRP}), number of grade-\Rmnum{3} level-A hospitals (\textit{3A.Hospital}), daily average  temperature (\textit{Temperature}) and 2 categorical variables: tier of the city (\textit{City.Tier}) and region of the city (\textit{Region}), which might be potentially important.
A detailed description of the above variables is summarized in Table \ref{t1}.
\begin{table}
\vspace{0.25cm}
\setlength{\belowcaptionskip}{0cm}
\caption{\label{t1} Description of variables used in the study. }
 \centering
\resizebox{\textwidth}{!}{\small{
\linespread{1.3}
\begin{tabular}{rl}
\toprule
Variable & \multicolumn{1}{c}{Description} \\\hline
$Sevendays.Cucase_i$ & Number of confirmed cases during the first seven days  of\\
& the outbreak in city $i$\\
$City.Tier_i$  \quad& Tier of city $i$ which has six levels, namely $1, \ldots, 6$\\
$Travel.Intensity_i$ \quad &Average travel intensity of 14 days prior to the date of\\
&the first confirmed case in city $i$ \nonumber\\
$\textit{Pop.2018}_i$  \quad  &Population size (in millions) in $\log$ scale of city $i$ in 2018\\
$Region_i$ \quad  &Region of city $i$ which has seven levels, namely North, East,\\
\quad  &South, Central,  Northwest, Southwest and Northeast\\
$Dis.WH_i$ \quad  &Distance in $\log 10$ scale between Wuhan and  city $i$ \\
$Total.Flow_i$ \quad  &Passenger volume (in millions) in $\log$ scale from Wuhan to city $i$ \\
&by various means of transportation during the whole of 2018 \\
$PGRP_i$ \quad  &Per capita gross regional product in 2018 of city $i$ \\
$3A.Hospital_i$  \quad &Number of grade-\Rmnum{3} level-A hospitals in city $i$ \\
$Arr.Time_i$  \quad &Date of the first confirmed case in city $i$, where 31 December \\
&2019 is day 0\\
$Temperature_i$  \quad&Average daily temperature of fourteen days prior to the date of \\
&the first confirmed case in city $i$ \nonumber\\
$K.Resp_i$ \quad&A binary variable indicating whether or not control measure $K$ \\
&was implemented in city $i$, $K.Resp_i=1$ if city $i$ has implemented \\
&$K$ prior to the outbreak\\
$K.Resp_i^{*}$ \quad&A binary variable indicating whether or not control measure  $K$\\
&was implemented in city $i$, $K.Resp_i^{*}=1$ if city $i$ has implemented \\
&$K$ before or during its first week of outbreak\\
$K.Date_i$ \quad&Number of days that control measure $K$ has been implemented $K$ \\
&prior to the outbreak in city $i$, if $K.Resp_i=0$, then $K.Date_i=0$\\
&For example, if the date of the first confirmed case in city $i$ is on 28  \\
&January 2020, and  $K$ was implemented in city $i$ on 25 January 2020,  \\
&then $K.Date_i=3$ \\
$K.Date_i^{*}$ \quad&Timing of implementing control measure K in city i, where 31  \\
&December 2019 is day 0, if $K.Resp_i^{*}=0$, then $K.Date_i^{*}=0$ \\
&For example, if the date of the first confirmed case in city $i$ is on 28  \\
&January 2020, then $K.Date_i=28$ \\\bottomrule
\end{tabular}}}
\vspace{-0.75cm}
\end{table}

As for the source of data about the explanatory variables in Table \ref{t1}, the travel intensity in each city (\textit{Travel.Intensity}) was obtained from the Baidu Migration Map (\url{http://qianxi.baidu.com}).
We calculated the average of daily temperatures in each city (\textit{Temperature}) over the first 14 days before the arrival of the first confirmed case from the China Weather website (\url{http://www.weather.com.cn}).
From the Baidu Baike (\url{https://baike.baidu.com}), we acquired the China's city-tier classification (\textit{City.Tier}) and the region (\textit{Region}) in which each city is located.
The latest data of  the per capital gross regional product (\textit{PGRP}) was from the China City Statistical Yearbook (\url{http://www.stats.gov.cn}).
The number of grade-\Rmnum{3} level-A hospitals (\textit{3A.Hospital}) in each city was collected from the China kang website (\url{http://www.cnkang.com}).
The sources of the data of the remaining 10 explanatory variables and the response variable in Table \ref{t1} are described in \cite{Tian:2020}.
\subsection{Modeling  the Effects of COVID-19 Control Measures}
\label{sec:Modeling}
\subsubsection{Linear Regression Model  With Categorical and Continuous Predictors}
\label{sec:lrm}
We apply linear regression model \eqref{linear1} to investigate the effects of the relevant factors on the spread of COVID-19. Here, $Y$ is the response variable $\log(\textit{Sevendays.Cucase})$, $X_{\mathcal{I}_{1}}=(X_{i,1},\ldots,X_{i,5})^{\mathrm{T}}$ and  $X_{\mathcal{I}_{2}}=(X_{i,1},\ldots,X_{i,6})^{\mathrm{T}}$ are used to represent the categorical variable $City.Tier$ and $Region$, respectively. The remaining explanatory variables are 3  binary variables: $Enter.Resp$, $Bus.Resp$, $Railway.Resp$ and 11 continuous explanatory variables: $Travel.Intensity$, $Pop.2018$, $Dis.WH$, $Total.Flow$, $PGRP$, $3A.Hospital$, $Arr.Time$, $Temperature$, $Enter.Date$, $Bus.Date$, $Railway.Date$.
We will apply VIBIM to analyze the data.
\subsubsection{Poisson Regression Model}
\label{sec:prm}
\cite{Tian:2020} constructed a
 Poisson regression model
 \begin{flalign}
&\mathrm{log}(\mathrm{E}[Y|\widetilde{X}_1,\ldots,\widetilde{X}_{10}])=\widetilde{\beta}_0+\widetilde{\beta}_1\widetilde{X}_1
+\widetilde{\beta}_2\widetilde{X}_2+\ldots +\widetilde{\beta}_{p}\widetilde{X}_{10} \label{Poisson model}
\end{flalign}
 of the response variable $Sevendays.Cucase$ on 10 explanatory variables: $Dis.WH$, $Arr.Time$, $Total.Flow$, $Pop.2018$, $Enter.Date^{*}$, $Enter.Resp^{*}$, $Bus.Date^{*}$, $Bus.Resp^{*}$, $Railway.Date^{*}$,  $Railway.Resp^{*}$, where
$Total.Flow$ and $Pop.2018$ are offset variables. In \eqref{Poisson model}, $Y$ is assumed to be a Poisson random variable given $\widetilde{X}_1,\ldots,\widetilde{X}_p$. Note that we use $\widetilde{X}$ and $\widetilde{\beta}$ to avoid confusion with the notation in \eqref{linear1}. For comparison purposes, we also consider \cite{Tian:2020}'s model and denote it  by ``Poisson" in the following.

\section{Analysis of the COVID-19 Data}
\label{sec:analysis}
In this section, we continue the analysis of  linear model \eqref{linear1} with the response  and the 16 explanatory variables described  in Table \ref{t1}. In subsection 5.1, three variable selection methods: group LASSO, group SCAD, and group MCP together with four importance measures SOIL \citep{Ye:Yi:Yang:2018}, LMG \citep{Lindeman:Merenda:Gold:1980}, RFI1 and RFI2 \citep{Breiman:2001} are used to select predictors related to $Sevendays.Cucase$. Subsection 5.2 presents the estimates for the resultant models from VIBIM as well as the three selection methods.
Comparisons of different models in terms of information criteria and prediction accuracy are provided in subsection 5.3. We assess the stabilities of the selected models and perform diagnostic analyses in subsection 5.4. A discussion of  an interesting collinearity issue is  provided in subsection 5.5.

\subsection{A Preliminary Analysis of Variable Selection and Variable Importance}
\label{sec:preanalysis}
We first apply the three penalized regression methods with the tuning parameters selected by the standard BIC. The selected sets of important explanatory variables from group MCP, group SCAD and group LASSO are abbreviated as follows:

\makeatletter\def\@captype{table}\makeatother
%\vspace{0cm}
\setlength{\belowcaptionskip}{0.25cm}
\resizebox{\textwidth}{!}{\small{
\linespread{1.3}
\begin{tabular}{rl}
%Notation   & \multicolumn{1}{c}{The selected set of important variables} \\\hline
 $\mathrm{M_{gMCP}}^{\dag}$ =&\textit{\{Dis.WH, 3A.Hospital, Total.Flow, Pop.2018, Enter.Date\}}\\
$\mathrm{M_{gSCAD}}^{\dag}$ =&\textit{\{Dis.WH, 3A.Hospital, Total.Flow, Pop.2018, Enter.Date, Region}, \\
       &\textit{Temperature, Railway.Resp\}}\\
$\mathrm{M_{gLASSO}}^{\dag}$ =& \textit{\{Dis.WH, 3A.Hospital, Total.Flow, Pop.2018, Enter.Date, PGRP}, \\
       &\textit{Temperature, Region\}}. \\
\end{tabular}}}
\vspace{0cm}
\noindent These models are labeled as ``$\mathrm{gMCP}^{\dag}$'', ``$\mathrm{gSCAD}^{\dag}$'', ``$\mathrm{gLASSO}^{\dag}$'' respectively.

As can be seen, the different methods produce different results, making it difficult to identify which variables are really important. In addition, as mentioned already, the stability of such results is a big issue. To get a good sense of instability of  the three methods, we calculate the size of the symmetric difference between the originally selected model and that based on the perturbed data and the reduced data, denoted by PIVS and SIVS, respectively; see \cite{Nan:Yang:2014} for the implementation details. In our procedure, we denote by $\tau$ the perturbation size in PIVS and let $\rho$ be the proportion of observations removed from the original dataset in SIVS. The  results based on 100 replications are
gathered in Table S14 in Section S.2 of the Supplementary Material to save the space. We can see that the instabilities of group LASSO and group SCAD are unacceptably high, especially group SCAD. For example, when only 5\% of the data are removed, the SIVS value of group SCAD  is 5.11, that is, group SCAD would choose a model with more than 5 variables different on average.

We now apply the VIBIM procedure in Section 2.2 to address the observed high instabilities of model selection methods.  Besides SOIL, we also consider several existing methods, namely LMG , RFI1 and RFI2.  For SOIL, we choose $\psi=1$ in \eqref{BIC_weight} for the BIC-p weighting. The importances of the top 10 variables measured by SOIL, LMG   , RFI1 and RFI2 are shown in Table \ref{real data}.
\begin{table}
\setlength{\belowcaptionskip}{0pt}
 \caption{\label{real data} Top 10 variables for different importance measures.}
  \centering
\resizebox{1\textwidth}{!}{\small{
\begin{tabular}{lcccccccccccc}
\toprule
&Rank & SOIL                  &        & LMG        &        & RFI1            &             & RFI2            &     \\\hline
&1    &Dis.WH           &1.000	 &Dis.WH           &0.281   &Total.Flow  	  &1.000          &Total.Flow  	  &1.000\\
&2    &3A.Hospital      &1.000	 &Region           &0.245    &Pop.2018   	     &0.753          &Dis.WH             &0.590\\
&3    &Total.Flow       &0.990	 &City.Tier        &0.101    &Region    	     &0.666               &Region            &0.516\\
&4    &Pop.2018         &0.883	 &Total.Flow       &0.090    &Dis.WH            &0.610        &Pop.2018        &0.443\\
&5    &Enter.Date       &0.048   &Pop.2018          &0.090  &City.Tier          &0.541           &Temperature    &0.319\\
&6    &PGRP            &0.017   &3A.Hospital   &0.055     &Temperature         &0.487       &City.Tier           &0.281\\
&7    &Temperature      &0.005	 &Arr.Time   &0.033      &3A.Hospital       &0.382        &PGRP                 &0.171\\
&8    &Railway.Resp     &9e-4	 &PGRP        &0.024          &PGRP            &0.280              &3A.Hospital       &0.164 \\
&9    &Region         &5e-9    &Temperature   &0.023          &Arr.Time        &0.164           &Travel.Intensity   &0.154 \\
&10   &Bus.Resp	  &4e-10   &Travel.Intensity  &0.017    &Travel.Intensity   &0.078       &Arr.Time            &0.127\\
\bottomrule
 \end{tabular}}}
\vspace{-0.5cm}
\end{table}

It can be seen that the SOIL importance values of \textit{Dis.WH}, \textit{3A.Hospital}, \textit{Total.Flow} and \textit{Pop.2018} are close to 1, and they are also in the top 10 for the LMG, RFI1 and RFI2. The importances of \textit{Dis.WH}, \textit{Total.Flow} and \textit{Pop.2018} are in line with our expectations. Given the fact that hospitals play an important part in the early detection of coronavirus disease, it is appropriate that \textit{3A.Hospital} is considered as an important variable.
 Moreover, \textit{Enter.Date} is regarded as a slightly important variable by SOIL, which indicates that the timing of closing entertainment venues  affected the number of confirmed cases to a certain degree.

 The importance value of \textit{Temperature} assigned by SOIL is only 0.005, which is coincident with \cite{Yao:2020}'s conclusion that there is no significant correlation between the cumulative number of confirmed cases and temperature, whereas RFI1 and RFI2 presented a different view. Also, the top 5 variables for SOIL are all included in $\mathrm{gMCP}^{\dag}$, $\mathrm{gSCAD}^{\dag}$, and $\mathrm{gLASSO}^{\dag}$. In addition, except for those produced by SOIL, the importance values of the other methods are not
very discriminative, making it difficult to compare the importance of variables. Overall, SOIL seems to be more informative. To save space, the  analysis results without interactions  are provided in Section S.2 of the Supplementary Material, including coefficient estimations, information criterion values, prediction accuracies, instability measures and regression diagnostics.

The results of diagnostic analyses for  no interaction models (see Table S19 in  the Supplementary Material)
show that the  residuals  exhibit autocorrelation, and there may be interactions between some predictors. Thus,  in addition to the 16 explanatory variables in Table \ref{t1}, we add the interactions between top 8 variables according to SOIL importances in  Table \ref{real data} whose  importance values are greater than the threshold value $c=$1e-4.  From now on, the set of  variables for constructing  candidate models is  given by the 16 main effects together with all possible interactions between the above 8 variables identified by SOIL, hence a total of $16+ \binom{8}{2}=44$  variables.

For comparison, we also apply penalized regression methods to the data with the added interactions. The selected sets of important explanatory variables with interaction from group MCP, group SCAD and group LASSO are  as follows:

\makeatletter\def\@captype{table}\makeatother
\vspace{0.3cm}
\setlength{\belowcaptionskip}{0.25cm}
%\caption{\label{models2}\linespread{0.5} The resultant log-linear models with interaction from several methods.}
\resizebox{\textwidth}{!}{\small{
\linespread{1.3}\selectfont
\begin{tabular}{rl}
 $\mathrm{M_{gMCP}}$ =& \textit{\{Pop.2018, Dis.WH, Total.Flow, Enter.Date, 3A.Hospital,}\\
 & \textit{3A.Hospital*Railway.Resp, PGRP*Temperature\}}\\
 $\mathrm{M_{gSCAD}}$ =&\textit{\{Dis.WH, Total.Flow, 3A.Hospital, Region, Dis.WH*Enter.Date, }\\
 & \textit{Total.Flow*Pop.2018, Total.Flow*PGRP, 3A.Hospital*Temperature, }\\ & \textit{3A.Hospital*Railway.Resp, Enter.Date*Railway.Resp, PGRP*Temperature\}}\\
$\mathrm{M_{gLASSO}}$ = &\textit{\{Pop.2018, Dis.WH, Total.Flow, 3A.Hospital, PGRP, City.Level,}\\ &\textit{Region, Dis.WH*Enter.Date, 3A.Hospital*Temperature}\\
&\textit{3A.Hospital*Railway.Resp, Pop.2018*PGRP, PGRP*Temperature\}}\\
\vspace{0cm}
\end{tabular}}}
\noindent These models  are labeled as ``$\mathrm{gMCP}$'', ``$\mathrm{gSCAD}$'' and ``$\mathrm{gLASSO}$'', respectively. It is seen that
the models are quite different. Considering that the interaction terms and their main effects are possibly highly correlated, group MCP, group SCAD and group LASSO may be unstable and fail to identify some important interaction terms.

With VIBIM, the top 12 variables in SOIL importance  are presented in Table \ref{Importance2}. It can be seen that the top 4 variables have the same rank and their importance values are close to those in Table \ref{real data}.
The main difference between SOIL importance values in Table \ref{real data} and Table \ref{Importance2} is the addition of 6 interactions: \textit{3A.Hospital*Railway.Resp, Pop.2018*PGRP, Dis.WH*Total.Flow, PGRP*Temperature, Dis.WH*Enter.Date, 3A.Hospital*Temperature}. It is also of interest to note that \textit{3A.Hospital*Railway.Resp} is the most important interaction although its main effect  \textit{Railway.Resp} is less important.

\begin{table}
\setlength{\belowcaptionskip}{1.5pt}
 \caption{\label{Importance2} Top 12 variables for SOIL  (after including interactions between top 8 variables).} \vspace{-0.5 cm}
 \centering
\resizebox{1\textwidth}{!}{\small{
\begin{tabular}{lcccccccccccc}
\toprule
&Rank & Variable         &Importance                     & Rank      & Variable         &Importance    \\\hline
&1    &Dis.WH           &1.000	 &2    &3A.Hospital &0.960  \\ &3 &Total.Flow  &0.914
&4    &Pop.2018  &0.731\\	  &5    &3A.Hospital*Railway.Resp  &0.055   &6 &Pop.2018*PGRP &0.051  \\
&7    &Dis.WH*Total.Flow &0.048	 &8    &PGRP*Temperature   &0.003\\ &9 &Dis.WH*Enter.Date &0.001
&10    &Enter.Date &0.001	\\ &11    &Bus.Resp &5e-5 &12 &3A.Hospital*Temperature &7e-08\\
\bottomrule
 \end{tabular}}}
\vspace{0.5cm}
\end{table}
We narrow down the sequence of  nested models by information criteria, standard AIC and BIC.  Denote by $\mathrm{SOIL_i}$ $(i=5,\ldots,9)$ the resultant linear models based on the first $i$ variables of \textit{\{Dis.WH, 3A.Hospital, Total.Flow, Pop.2018, 3A.Hospital*Railway.Resp, Pop.2018*PGRP, Dis.WH*Total.Flow, PGRP*Temperature, Dis.WH*Enter.Date\}}, then the minimizers of BIC and AIC are $\mathrm{SOIL_5}$ and $\mathrm{SOIL_9}$, respectively. Thus, for VIBIM, the most plausible models
are  $\{\mathrm{SOIL_5},\ldots,\mathrm{SOIL_9}\}$.
Table \ref{BIC2} summarizes AIC and BIC values of various regression models. We can see that no model can be optimal in aspects of both AIC and BIC.  The overall performances of $\mathrm{SOIL_8}$, $\mathrm{SOIL_9}$ and $\mathrm{gMCP}$  seem comparable  in terms of the two information criteria,  although significant differences will be seen later.

\begin{table}
\setlength{\belowcaptionskip}{1.5pt}
\caption{\label{BIC2} AIC and BIC values of various regression models.} \vspace{-0.15cm}
\linespread{1.25} \vspace{-0.5 cm}
 \centering
\resizebox{1\textwidth}{!}{\small{
\begin{tabular}{llcccccccccccccc}
\toprule
&               &$\mathrm{SOIL}_5$  &$\mathrm{SOIL}_6$    &$\mathrm{SOIL}_7$       &$\mathrm{SOIL}_8$    &$\mathrm{SOIL}_9$  &$\mathrm{gLASSO}$   &$\mathrm{gSCAD}$    &$\mathrm{gMCP}$   \\ \hline
&$\mathrm{AIC} $  &684.278   &685.849   &684.501    &674.953   &671.371   &672.426   &655.116   &674.940\\
&$\mathrm{BIC}$    &710.110   &715.372   &717.714   &711.856   &711.965    &757.305   &725.232   &708.153 \\
\bottomrule
\end{tabular}}}
\vspace{0.5cm}
\end{table}
\subsection{Model Estimation and Results}
\label{sec:estimation}
The corresponding coefficient estimates of above models are reported in Table \ref{coefficients2}. As can be seen, \textit{Dis.WH}, \textit{3A.Hospital}, \textit{Pop.2018} and \textit{Total.Flow} are highly significant and the signs of the estimated coefficients in accordance with our prior expectations. In addition, the coefficients of 5 interactions in $\mathrm{SOIL}_9$ are significant at the significance level 0.05 except \textit{Pop.2018*PGRP}. We know from the fact that the coefficient on \textit{Dis.WH*Enter.Date} is negative, i.e., \textit{Enter.Date} has a negative effect (note that \textit{Dis.WH} takes only positive values), which provides supporting evidence that the
longer timing of closing entertainment venues before the date of the first confirmed case, the fewer number of confirmed cases. It is notable that the marginal effect of \textit{3A.Hospital} on the response variable varies with the presence or absence of \textit{Railway.Resp}. For instance, in $\mathrm{SOIL}_9$, when $\textit{Railway.Resp}=0$, the effect of \textit{3A.Hospital} is 0.032; when $\textit{Railway.Resp}=1$, the effect of \textit{3A.Hospital} is $0.032+0.141=0.173$. Perhaps this is because grade-\Rmnum{3} level-A hospitals in cities where the railway was banned had higher daily nucleic acid testing capability.

For $\mathrm{SOIL_i}$ $(i=5,\ldots,9)$, any two nested models of them coincide, that is, the signs of estimated coefficients of the smaller model are exactly the same as those in the larger model, with only one exception of \textit{Pop.2018*PGRP}. This is reasonable because \textit{Pop.2018*PGRP} is insignificant in $\mathrm{SOIL_i}$ $(i=6,\ldots,9)$ and therefore its sign may be unstable in different nested models. Therefore, our procedure offers  reliable insights of the relationship between $Sevendays.Cucase$ and important predictors based on multiple models. In contrast, group LASSO and group SCAD select too many
variables that are not significant. Also, group MCP possibly excludes at least one  of the relevant variables, such as \textit{Dis.WH*Total.Flow}. To verify the usefulness of this term, we fit the model by adding \textit{Dis.WH*Total.Flow} to $\mathrm{gMCP}$. The result shows that \textit{Dis.WH*Total.Flow} is highly significant ($p$-value is 0.008), suggesting that group MCP may be overly parsimonious.
\begin{landscape}
\begin{table}
\caption{ \label{coefficients2}Coefficients ($p$-values) of the variables of various regression models, and ``--" means that the variable does not exist in models.}
\centering
 \linespread{1.25}
\resizebox{0.95\linewidth}{!}{\footnotesize{
\begin{tabular}{lccccccHccc}
\toprule
&               &$\mathrm{SOIL}_5$         &$\mathrm{SOIL}_6$  &$\mathrm{SOIL}_7$ &$\mathrm{SOIL}_8$  &$\mathrm{SOIL}_9$  &$\mathrm{SOIL}_{10}$   &gLASSO &gSCAD  &gMCP\\ \hline
&Dis.WH      &-2.259(2e-16)     &-2.245(2e-16)  &-2.502(2e-16)   &-2.573(2e-16)       &-2.575(2e-16)
&-2.554(2e-16)    &-1.569(8e-9) &1.535(7e-9)  &-2.213(2e-16)\\
&3A.Hospital  &0.032(2e-6)       &0.028(0.002)     &0.029(0.002)      &0.0328(3e-4)    &0.032(4e-4)
&0.032(5e-4) &0.038(0.001)  &0.034(8e-8)   &0.029(1e-5)\\
&Total.Flow      &0.066(5e-6)     &0.066(5e-6)   &0.299(0.022)    &0.428(0.001)     &0.411(0.002)
&0.413(0.002) &0.038(0.024)   &0.143(3e-7)  &0.058(5e-5)\\
&Pop.2018       &0.251(8e-5)       &0.219(0.007)     &0.232(0.004)      &0.329(1e-4)   &0.319(2e-4)
&0.324(2e-4)    &0.314(0.031)  &--  &0.270(2e-5) \\
&3A.Hospital*Railway.Resp  &0.135(0.014)   &0.138(0.013)  &0.117(0.037)  &0.1244(0.025)     &0.141(0.011) &0.138(0.013)  &0.134(0.018)   &0.189(0.001)  &0.170(0.002)      \\
&Pop.2018*PGRP        &--   &0.008(0.518)          &0.009(0.442)        &-0.003(0.835)   &-0.002(0.864)          &-0.002(0.868)  &0.025(0.299) &--  &-- \\
&Dis.WH*Total.Flow    &--    &--     &-0.086(0.071)  &-0.137(0.006)    &-0.131(0.008) &-0.132(0.007)  &--  &-- &-- \\
&PGRP*Temperature  &--        &--         &--   &0.003(0.001)  &0.003(7e-4) &0.003(7e-4) &0.001(0.275)  &0.001(0.001)   &0.002(0.006)\\
&Dis.WH*Enter.Date   &--      &--     &--      &-- &-0.039(0.020)    &-0.130(0.530) &-0.046(0.009) &-0.046(0.016) &--\\
&Enter.Date                   &--       &--    &--  &--     &-- &0.263(0.660) &--  &--  &-0.119(0.016) \\
&PGRP                    &--       &--    &--  &--     &-- &--       &0.032(0.309)  &-- &-- \\
&City.Tier1    &--       &--            &--            &--          &--     &--  &-0.091(0.914)   &-- &-- \\
&City.Tier2      &--       &--          &--      &--              &--         &--     &0.346(0.392) &--   &-- \\
&City.Tier3     &--       &--          &--      &--              &--         &--   &0.223(0.406) &--  &-- \\
&City.Tier4   &--       &--          &--      &--              &--         &-- &0.267(0.113)    &-- &-- \\
&City.Tier5    &--       &--          &--      &--              &--         &--     &0.251(0.055)  &--    &--   \\
&Region.North   &--       &--          &--      &--              &--         &--         &-0.167(0.424)  &0.138(0.493)  &-- \\
&Region.East    &--       &--          &--      &--              &--         &--   &0.097(0.702)  &0.110(0.653)  &-- \\
&Region.South      &--       &--          &--      &--              &--         &--   &-0.030(0.909)  &0.032(0.897) &--  \\
&Region.Central   &--       &--          &--      &--              &--         &--    &0.723(0.008) &0.724(0.006)  &--  \\
&Region.Northwest    &--       &--          &--      &--              &--         &--     &-0.048(0.818) &0.027(0.895)  &-- \\
&Region.Southwest   &--       &--          &--      &--              &--         &--  &0.012(0.956)  &0.053(0.796) &-- \\
&3A.Hospital*Temperature &--       &--          &--      &--              &--         &--  &9e-4(0.252) &9e-4(0.193) &-- \\
&Bus.Resp  &--       &--          &--      &--              &--         &--  &--  &0.382(0.085)  &-- \\
&Total.Flow*Pop.2018  &--       &--          &--      &--              &--         &--  &--  &-0.056(4e-7)  &--  \\
&Total.Flow*PGRP  &--       &--          &--      &--              &--         &--  &--   &-0.004(0.152) &--   \\
&Enter.Date*Railway.Resp  &--       &--          &--      &--              &--         &--  &-- &-0.231(0.044) &-- \\
\bottomrule
 \end{tabular}}}
\end{table}
\end{landscape}
\subsection{Model Comparison}
\label{sec:comparison}
We compare the performances of various models in terms of  prediction accuracy (i.e., mean square error (MSE))
using cross-validation. The COVID-19 dataset is randomly split into a training set with proportion of 4/5 and a test set with proportion of 1/5. On the training set, we select important variables and estimate the coefficients of the model by applying the least squares method.
The test set is used to evaluate the prediction performance of the models built on the training set. Table \ref{MSE2} reports the average results over 1000 runs of MSE of prediction on the test set, as measured in the original scale of the response
variable (i.e., for the linear regression models, the predictions are exponentiated and compared to the true non-transformed response values).

\begin{table}
\renewcommand\arraystretch{1.25}
\vspace{-0.2cm}
\caption{ Prediction errors of various regression models.}\label{MSE2}
\centering
\begin{threeparttable}
\vspace{0cm}
\begin{tabular}{llcccccccccc}
\toprule
&           &Poisson &$\mathrm{SOIL}_5$    &$\mathrm{SOIL}_6$      &$\mathrm{SOIL}_7$      &$\mathrm{SOIL}_8$          \\ \hline
&MSE &6600.871 &885.373   &825.305    &754.367   &741.382  \\
& &(356.499) &(24.047)   &(24.664)   &(19.081) &(18.896) \\ \hline
& &$\mathrm{SOIL}_9$        &$\mathrm{gLASSO}$   &$\mathrm{gSCAD}    $    &$\mathrm{gMCP}$  \\  \hline
&MSE  &736.289   &960.775  &900.339  &$890.761^{\dag}$   \\
&    &(18.422)  &(24.344)  &(26.537)   &$(24.172^{\dag})$  \\\bottomrule
\end{tabular}
\begin{tablenotes}
\item Note: The standard errors are given in parenthesis.
\item $\dag$ The results for gMCP are obtained by removing overly large outliers (greater  than $10^ {5}$) over 1000 runs.
\end{tablenotes}
\end{threeparttable}
\vspace{0cm}
\end{table}
The results in Table \ref{MSE2} indicate that the proposed procedure yields more accurate predictions than  group LASSO, group SCAD and group MCP.  In addition, with the same number of variables, $\mathrm{SOIL_7}$ yields more accurate predictions than group MCP.  In contrast, the prediction accuracies of Poisson are quite poor, that is, the MSE of Poisson is one magnitude bigger than that of model $\mathrm{SOIL_i}$ ($i=7,8,9$). It is also worth pointing out that, compared to the results of no interaction models (see Table S18 in Section S.2 of the Supplementary Material), taking interaction into account significantly improves the prediction accuracy.
\subsection{Model Stability and Diagnostic}
 \label{sec:diagnostic}
In this subsection, we examine the stabilities of the selected models. Also, diagnostic analyses are carried out to check whether model assumptions have been violated.

We use the PIVS and SIVS measures introduced in Section 5.1 to evaluate the instability of various methods. The PIVS and SIVS results  in Table \ref{stabilityinteraction} show that $\mathrm{SOIL}$ outperforms group MCP, group SCAD and group LASSO across all settings. More specifically, both PIVS and SIVS values of group MCP, group LASSO and group SCAD are typically greater than 3, 4, 8, respectively. In contrast, in nearly all of the settings we consider, our method  would choose a model with  no more than 2 variables different on average.
\begin{table}
\setlength{\belowcaptionskip}{10pt}
\caption{ PIVS and SIVS of various methods.}\label{stabilityinteraction}
\centering
\begin{threeparttable}
\small{
\begin{tabular}{lcccccccccccc}
\toprule
&Method          &\multicolumn{2}{c}{PIVS}       &\multicolumn{2}{c}{SIVS}   \\
&             &$\tau=0.1$ &$\tau=0.2$       &$\rho=0.05$   &$\rho=0.1$  \\ \hline
&$\mathrm{SOIL}_5$         &1.66           &1.86               &1.78    &2.04                        \\
&$\mathrm{SOIL}_6$         &1.76           &2                 &1.92       &2.06                       \\
&$\mathrm{SOIL}_7$         &1.58           &1.82             &1.8      &1.72                         \\
&$\mathrm{SOIL}_8$         &1.2           &1.56                &1.86    &2.06                        \\
&$\mathrm{SOIL}_9$         &0.64           &1.18               &1.52    &1.86                         \\
&group LASSO         &3.96          &4.03           &5.03     &6.58                        \\
&group SCAD          &8.4           &8.8                &9.82   &9.93                             \\
&group MCP           &3.49           &3.11              &3.23    &2.69                         \\
\bottomrule
\end{tabular}}
   \begin{tablenotes}
     \item Note: We use the  averaged size of the symmetric difference between the originally top $i$ variables and that based on the modified version of the data to measure the instability of our procedure.
   \end{tablenotes}
  \end{threeparttable}
\end{table}

In order to verify the assumptions of the models, we perform a series of regression diagnostics for plausible candidate models: $\mathrm{SOIL_i}$ $(i=5,\ldots,9)$, in regard to model significance, heteroscedasticity, normality, multicollinearity and error independence. For testing these five problems, we use $F$ test, Breusch-Pagan test, Shapiro-Wilk test, Variance Inflation Factor (VIF) and Durbin-Watson test, respectively. If the VIF of a variable is greater than 4, we consider that the variable has a collinearity problem with other variables. The detailed results are reported in Table \ref{diagnostics2}, and ``$\checkmark$'' indicates that a specified model has the given property.
\begin{table}\setlength{\belowcaptionskip}{10.5pt}
 \caption{Results of diagnostic analyses for $\mathrm{SOIL_i}$, $i=4,\ldots,7$.}\label{diagnostics2}
 \centering
\begin{threeparttable}
\small{
\begin{tabular}{lccccccHHHHHH}
\toprule
&Property                            &$\mathrm{SOIL}_5 $           &$\mathrm{SOIL}_6  $    &$\mathrm{SOIL}_7  $  &$\mathrm{SOIL}_8$ &$\mathrm{SOIL}_9$ &$\mathrm{SOIL}_{10}$    \\\hline
&Model significance       &$\checkmark$($\checkmark$)&$\checkmark$($\checkmark$)  &$\checkmark$($\checkmark$)   &$\checkmark$($\checkmark$)  &$\checkmark$($\checkmark$) &$\checkmark$($\checkmark$)    \\
&Homoscedasticity         &$\checkmark$($\checkmark$)&$\checkmark$($\checkmark$)  &$\checkmark$($\checkmark$)    &$\checkmark$($\checkmark$)  &$\checkmark$($\checkmark$) &$\checkmark$($\checkmark$)  \\
&Normality                &$\checkmark$($\checkmark$)&$\checkmark$($\checkmark$)&$\checkmark$($\checkmark$)    &$\checkmark$($\checkmark$)  &$\checkmark$($\checkmark$) &$\checkmark$($\checkmark$) \\
&Error independence       &{\scriptsize{\XSolidBrush}}({\scriptsize{\XSolidBrush}})&{\scriptsize{\XSolidBrush}}({$\checkmark$})
&{\scriptsize{\XSolidBrush}}({$\checkmark$})    &{$\checkmark$}({$\checkmark$})   &{$\checkmark$}({$\checkmark$})&{$\checkmark$}({$\checkmark$})      \\
&No multicollinearity     &$\checkmark$              &\scriptsize{\XSolidBrush}              &\scriptsize{\XSolidBrush} &\scriptsize{\XSolidBrush} &\scriptsize{\XSolidBrush} &\scriptsize{\XSolidBrush} \\
\bottomrule
\end{tabular}}
   \begin{tablenotes}
     \item Note: The significance level $\alpha=0.05$ and the results of the case $\alpha=0.01$ are given in parenthesis.
   \end{tablenotes}
  \end{threeparttable}
\vspace{-0.2cm}
\end{table}
It can be seen that compared to no interaction models (see Table S19 in Section S.2 of the Supplementary Material), $\mathrm{SOIL_i}$ $(i=5,\ldots, 9)$ all have desirable properties: model significance, homoscedasticity and normality. Moreover, the results show no evident correlation between the
residuals of $\mathrm{SOIL_8}$ and $\mathrm{SOIL_9}$ at the significance level 0.05.

Through the comparison among models from variable selection methods and the proposed procedure in different aspects, including coefficient estimations, information criterion values, prediction accuracies and instability measures, we proclaim that our procedure outperforms the group variable selection methods.
\subsection{Discussion of Collinearity}
\label{sec:collinearity}
As we can see from Table \ref{diagnostics2}, there exists collinearity among some variables in $\mathrm{SOIL_i} $ $(i=6,\ldots, 9)$. Further examination shows that the collinearity issue lies with \textit{Total.Flow} and \textit{Dis.WH*Total.Flow} (the Pearson correlation coefficient between them is 0.990). Nonetheless, from Table \ref{coefficients2} we can see that the above two variables are significant for models $\mathrm{SOIL_8}$ and $\mathrm{SOIL_9}$  at the significance level 0.05 (the $p$-value of \textit{Dis.WH*Total.Flow} in $\mathrm{SOIL_9}$ is 0.007 after netting out the effect of others explanatory variables, see Section S.4 of the Supplementary Material for more details). In this section, we consider $\mathrm{SOIL}_9$ as an example to show the reliability of the inclusion of \textit{Dis.WH*Total.Flow}.

\begin{figure}
\vspace{0cm}
\centering
    \includegraphics[width=0.5\linewidth]{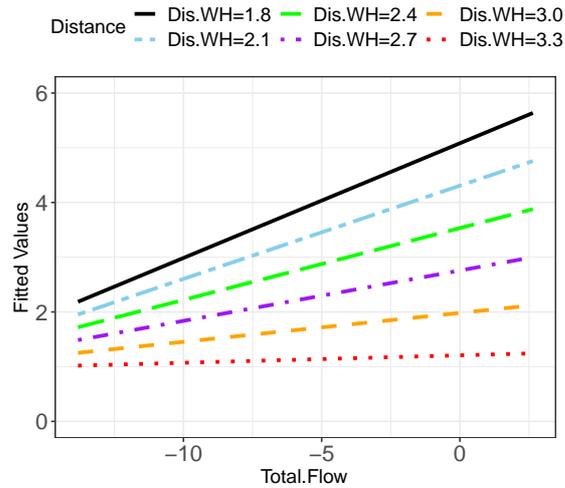}
    \caption{\label{Totalflow}Relationship between \textit{Total.Flow} and the fitted values given different values of \textit{Dis.WH}.}
  \vspace{-0.5cm}
\end{figure}

The marginal effect of  \textit{Total.Flow} on log(\textit{Sevendays.Cucase}) across a substantively meaningful range of \textit{Dis.WH} can be depicted graphically in Figure \ref{Totalflow}. We can clearly see that the larger value of \textit{Dis.WH}, the smaller slope of the line. In other words,  an increased \textit{Total.Flow} yields a higher increase in the number of confirmed cases for smaller \textit{Dis.WH}. In fact, cities far away from Wuhan typically have a lower population density, which may have led to a less within-population contact, thereby exhibiting a slower speed of growth in cases. Therefore, the sign of the interaction term seems to be a sensible representation of reality to some extent.

When faced with severe collinearity,  the ``standard"  approach is to drop one of the collinear variables. However, omitting an important variable generally results in biased (and inconsistent) estimators of the coefficients \citep{Wooldridge:2015:introductory}. For instance, when we drop \textit{Dis.WH*Total.Flow} from $\mathrm{SOIL}_7$, we obtain $\mathrm{SOIL}_6$, whose estimated coefficient of \textit{Dis.WH} and \textit{Total.Flow} are dramatically different from those of $\mathrm{SOIL}_7$. In cases of high collinearity, one worries about possibly much inflated standard errors of the corresponding coefficient estimates and it is also likely to encounter that the estimators and their significances can be unstable to small changes in the data. Hence, we conduct further analysis of the credibility and stability of the results in Table \ref{coefficients2} and significance of  \textit{Total.Flow} and \textit{Dis.WH*Total.Flow}.

We perform a guided simulation study to check credibility of the inclusion of \textit{Dis.WH*} \textit{Total.Flow}. To be specific, we  generate the simulation data by model $\mathrm{SOIL}_8$, i.e.,
\begin{flalign}
 \widetilde{Y} &= \widehat{\beta_0} + \widehat{\beta_1} Dis.WH+ \widehat{\beta_2} 3A.Hospital+ \widehat{\beta_3} Total.Flow +\widehat{\beta_4}Dis.WH*Total.Flow  \nonumber \\
 &+\widehat{\beta_5} Pop.2018+\widehat{\beta_6} 3A.Hospital*Railway.Resp  +\widehat{\beta_7}Pop.2018*PGRP  \nonumber \\ &+\widehat{\beta_8}PGRP*Temperature+\epsilon \label{SOIL7},
\end{flalign}
where $\widehat{\beta}_i$ is the estimated coefficients of model $\mathrm{SOIL}_8$, $\epsilon$ is drawn from normal distribution $N(0,\hat{\sigma}^{2})$,  where $\hat{\sigma}^2$ is the estimated error variance of $\mathrm{SOIL}_8$. Then we apply the VIBIM procedure to the simulated dataset and obtain top 8 predictors $\mathrm{SOIL}_8^{*}$, the threshold value $c$ in \textit {Step} 2 of VIBIM procedure is set as 1e-4. Also, we  generate the simulation data by model $\mathrm{SOIL}_{8-1}$, that is, variables in $\mathrm{SOIL}_{8}$ with  \textit{Dis.WH*Total.Flow} excluded and
obtain top the 7 predictors $\mathrm{SOIL}_{8-1}^{*}$ by VIBIM on the simulated dataset for comparison.
We record the number of times that \textit{Dis.WH},  \textit{Total.Flow} and \textit{Dis.WH*Total.Flow} are all included in $\mathrm{SOIL}_8^{*}\, (\mathrm{SOIL}_{8-1})$ and all significant at the significance level 0.05  over 1000  replications. We also apply a similar procedure to $\mathrm{SOIL}_9$. The results  are summarized in Table \ref{guide}. It can be seen that when \textit{Dis.WH*Total.Flow} is included in the true data generating model, the VIBIM procedure can correctly identify this interaction and its parent effects around 30\% of times. In contrast, if \textit{Dis.WH*Total.Flow} is excluded from the true data generating model, VIBIM rarely produces models such that \textit{Dis.WH},  \textit{Total.Flow} and \textit{Dis.WH*Total.Flow} are all included and all significant. The percentage difference provides a support to the VIBIM finding on the interaction term.
\begin{table}
\renewcommand\arraystretch{1.25}
\setlength{\belowcaptionskip}{10.5pt}
\caption{\label{guide} Results for guided simulation.}
\vspace{0cm}
\begin{tabular}{llccccccccccc}
\toprule
&            &$\mathrm{SOIL}_8$    &$\mathrm{SOIL}_{8-1}$      &$\mathrm{SOIL}_{9}$      &$\mathrm{SOIL}_{9-1}$     \\ \hline
&Times  &318 &10 &252 &4  \\
\bottomrule
\end{tabular}
\vspace{-0.2cm}
\end{table}

We have also performed an instability analysis on the issue (see Section S.3 of the Supplementary Material).  The results show that the instabilities of the significance of \textit{Dis.WH*Total.Flow} and \textit{Total.Flow} are really low. Also, it is instructive to point out that the partial correlation between \textit{Dis.WH*Total.Flow} and \textit{Total.Flow} in $\mathrm{SOIL}_9$ is $-0.638$. Thus, there is a strong reason to proclaim that the model $\mathrm{SOIL}_9$ with interaction term is a  stable and accurate description of the relationship between  log(\textit{Sevendays.Cucase}) and relevant predictors.

\section{Conclusion}
\label{sec:conclusion}
In this article, we consider the problem of  learning pairwise interactions in a linear regression model with both continuous and categorical predictors. To overcome the  weaknesses of existing approaches, we proposed a new interaction selection procedure (VIBIM) that delivers multiple strong candidate models with high stability and  interpretability. Our numerical results suggest that VIBIM gives superior performance for high-dimensional data. The VIBIM procedure  is applied to analyze the COVID-19 data. In light that some variables related to the control measures in \cite{Tian:2020} have possible causality flaws, we properly modified the questionable variables, and also added some potentially important variables to the COVID-19 dataset. We found a newly added important explanatory variable \textit{3A.Hospital} and 5 relatively important interaction terms \textit{3A.Hospital*Railway.Resp, Pop.2018*PGRP, Dis.WH*Total.Flow, PGRP*Temperature, Dis.WH*Enter}. Unlike the ``standard" approach of dealing with collinearity, we actually include some collinear variables, and this is supported by strong evidence. Moreover, in contrast to the coefficient estimates reported in \cite{Tian:2020}, the distance variable in models $\mathrm{SOIL_i}$ $(i=5,\ldots,10)$ are all of the expected sign.  A series of  analysis results demonstrate that our proposed procedure can lead to better models in terms of interpretability,  stability, reliability and accurate prediction than commonly used alternative methods.

\begin{center}
{\large\bf SUPPLEMENTARY MATERIAL}
\end{center}

\begin{description}

\item[Text document:] Supplementary Material for “Variable Importance Based Interaction Modeling with an
Application on Initial Spread of COVID-19 in China”. (.pdf file)

\item[R source code for VIBIM:] R source code to perform analysis described in the article. (.zip file)

\item[COVID-19 dataset:] Dataset used in the illustration of the VIBIM method in Sections 4-5. (.csv file)

\end{description}

\bibliographystyle{rss}
\bibliography{ref}
\end{document}